# MECHANISM OF HEATING THE SOLAR CORONA IN THE SPLITTING OF MASSIVE PHOTON PAIRS

I. K. Mirzoeva[a] *, and S. G. Chefranov[b]

[a] *Space Research Institute, Russian Academy of Sciences, Moscow, 117997 Russia*

[b] *Obukhov Institute of Atmospheric Physics, Russian Academy of Sciences, Moscow, 119017 Russia*

*e-mail: colombo2006@mail.ru

**Abstract**—Data obtained in the framework of the *INTERBALL-Tail Probe* (1995–2000) and *RHESSI* (from 2002 to the present) projects have revealed variations in the X-ray intensity of the solar corona in the photon energy range of 2–15 keV during the period of the quiet Sun. Previously, a hypothesis was proposed that this phenomenon could be associated with the effect of coronal heating. In the present study, a new mechanism of coronal plasma heating is proposed on the basis of the experimental data and the quantum theory of photon pairs that are produced from vacuum in the course of the Universe's expansion. A similar mechanism based on the splitting of photon pairs in the interplanetary and intergalactic space is also proposed to explain the observed microwave background radiation.

## 1. INTRODUCTION

In the previous studies [1–3], variations in the intensity of solar X-ray emission in the photon energy range from 2 to 15 keV during the period of the quiet Sun were investigated. The phenomenon of a decrease in the intensity of solar emission in narrow subranges of the X-ray spectrum in the photon energy range of 2–15 keV was revealed in 2005 [1] by analyzing the data of the *INTERBALL-Tail Probe* project. Further, this phenomenon was confirmed by the data of the *RHESSI* project. In [2, 3], the total range of soft X-ray (SXR) emission from 3 to 11 keV was divided into narrow subranges of width 1 keV, i.e., the X-ray spectrum was separately analyzed in the following subranges: 3–4, 4–5, 5–6, 6–7, 7–8, 8–9, 9–10, and 10–11 keV. Such a partition of the spectrum made it possible to observe a drop (in some cases, an increase) in the X-ray intensity of microflares and thermal background radiation of the solar corona. In some cases, the maximum drop in the X-ray intensity was shifted toward harder spectral range. Detailed analysis of the observational data revealed the subranges in which the X-ray intensity most often decreased: these are the subranges of 3–4 and 4–5 keV and, in some cases, the subranges of 7–8 and 8–9 keV. Along with this drop, an increase in the X-ray intensity was also observed in the spectral subrange of 10–11 keV compared to the other subranges in the total spectral range of 3–16 keV.

Let us consider an example of an atypical energy spectrum of the thermal X-ray background of the solar corona [3]. Figure 1 shows almost quiet thermal X-ray background of the solar corona recorded during 7 min on March 10, 2003. In the figure, one can see the above-mentioned drop in the X-ray intensity in the subrange of 3–4 keV and, simultaneously, an increase in the intensity in the subrange of 10–11 keV. This phenomenon is well illustrated in Fig. 2, which shows the energy spectrum of the thermal background of the solar corona measured in time interval from 13:44 to 13:51 UT on March 10, 2003. The photon energy (in keV) is plotted on the abscissa, and the intensities of each component in the energy range from 3 to 11 keV (in counts per second) are plotted on the ordinate. The plot is compiled from the data presented in Fig. 1. A typical descending X-ray spectrum in the photon energy range from 4 to 10 keV is disturbed by a sharp drop in the X-ray intensity in the subrange of 3–4 keV and by an increase in the subrange of 10–11 keV.

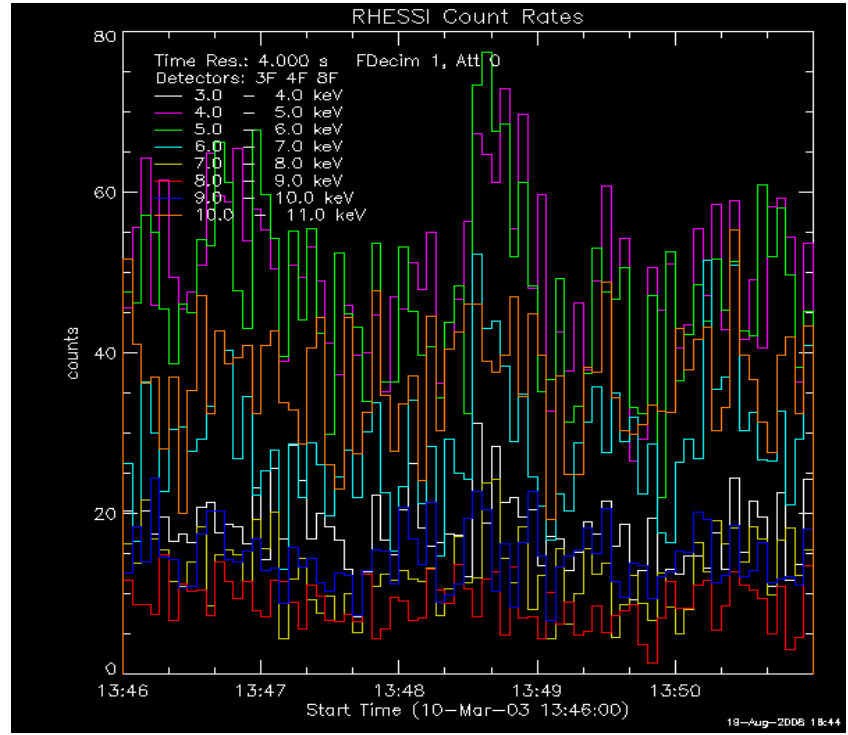

**Fig. 1.** Data from the *RHESSI* experiment recorded from 13:44 to 13:51 UT on March 10, 2003, with partition into 1-keV subranges

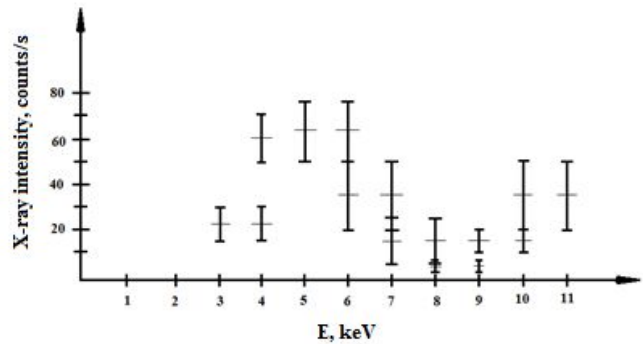

**Fig. 2.** Energy spectrum of the X-ray thermal background of the solar corona measured in the time interval from 13:44 to 13:51 UT on March 10, 2003.

It was ascertained in [2, 3] that such behavior of the X-ray spectrum was independent of the solar flares and other events that took place in the active flare regions of the solar corona. It was also supposed that the physical mechanism for the observed phenomena was associated with quantum processes and, in turn, this mechanism could be responsible for the anomalous heating of the solar corona plasma. For better understanding of what follows, let us make some digressions. In this paper, we will repeatedly use the following concepts and terms: axions, the Primakov effect (direct and inverse), conversion of axions, etc. Let us briefly explain their meanings.

The most challenging problem of modern astrophysics is the search for dark matter. As one of the candidates for the elementary particles that form dark matter, axions are considered. They are hypothetical electrically neutral pseudoscalar elementary particles that can be involved in the electromagnetic and gravitational interactions. In the original theory, the rest mass of an axion (in energy units) was assumed to be fairly large (100 keV). Later, it was suggested that the axion mass should be as low as 0.02 eV and this became the main difference of axions from the so-called "weakly interacting massive particles" (WIMPs), which are another candidate for the elementary particles forming dark matter. All particles that are similar to axions and unified by the general term "weakly interacting slim particles" (WISPs) are assumed to have a very low mass and weakly interact with other particles [4]. Initially, axions were introduced rather artificially to explain the violation of the CP-symmetry in strong interactions [5]. Theoretically, under the effect of a static electric or magnetic field, axion should spontaneously decay into two photons. This effect was called the inverse Primakov effect. Accordingly, the direct Primakov effect is the resonance conversion of photon into a pseudoscalar particle (axion) in the electric or magnetic field. Such conversions of axions into photons and back are often called the conversion of axions. Figure 3 shows the Feynman diagram that illustrates the conversion of an axion into a pair of photons and bach, i.e., the inverse and direct Primakov effects. In theory, due to the conversion of axions into photons, "vacuum" should acquire certain optical properties (birefringence and dispersion) in the magnetic or electric field. Many experiments on the search of axions are based on the Primakov effect (direct and/or inverse). It is proposed to consider the cores of stars (in particular, the Sun's core) to be sources of axions. In theory, there are a number of processes that should contribute to the formation of the axion flux from the Sun: the Primakov effect, Compton scattering, scattering of electrons by electrons and ions, recombination, and deexcitation.

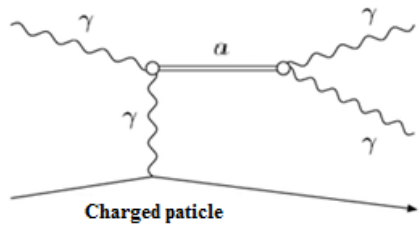

**Fig. 3.** Feynman diagram illustrating the conversion of an axion (*a*) into a pair of photons (γ) and back in the field of a charged particle (the inverse and direct Primakov effects).

To detect solar axions, a number of experiments were performed. In particular, the CAST (CERN Axion Solar Telescope) experiment is being conducted in CERN since 2003. The CAST experiment is aimed at detecting axions that are assumed to be emitted (due to the Primakov effect) by the solar core plasma with a temperature of ~15 × $10^6$ K [6]. No axions or expected effects that could confirm their existence have been observed in any experiment [7–9].

However, in 2014, J. Fraser et al. [10], scientists from the XMM-Newton International Space Observatory, reported on the indirect data that evidenced in favor of the existence of axions. The main idea of the XMM-Newton experiment was to attempt detecting solar axions that, in accordance with the Primakov effect, should convert into photons in the Earth's magnetic field. Of course, it is possible to detect only these resulting X-ray photons, which are a kind of marker indicating the presence of axions. The authors of the experiment believed that, in order to see these photons, the telescope should not be directed straight to the Sun, because several conversions have time to occur in the Earth's magnetosphere. Due to the motion of the Earth around the Sun, the mutual orientation of the geomagnetic field lines, the direction of observations, and the direction to the Sun changes throughout the year. Therefore, the telescope should detect seasonal variations in the X-ray background in the Earth's magnetosphere. The background associated with distant galaxies, interstellar gas, and cosmic radiation passing through the solar system should remain approximately constant, while the part of the background associated with axions should vary seasonably. In addition, for the purity of the experiment, the brightest X-ray sources (such as stars, star clusters, and other compact X-ray objects) were excluded from observations. Thus, a "clean X-ray sky" was obtained. Against this "cleaned" background, the desired seasonal variations in the X-ray background in the Earth's magnetosphere were observed. On the basis of the XMM-Newton experiment, it was concluded [10] that axions (the candidates for the dark matter particles) are actually produced in the Sun's core and converted into SXR emission in the Earth's magnetic field, which leads to seasonal variations in the X-ray background in the spectral range of 2–6 keV. Additionally, narrow axion lines associated with silicon, copper, and iron were detected in the same energy range.

Let us now return to the previously discussed results. In [2, 3], atypical behavior of the energy spectrum of the thermal X-ray background of the solar corona was analyzed. In narrow bands of the energy spectrum, drops and, in some cases, rises of the X-ray intensity were observed. Recall that we speak about the same photon energy range of 2–6 keV as in the XMM-Newton experiment. The difference is that variations in the X-ray intensity were found in the solar corona, whereas in the XMM-Newton experiment, the X-ray background of the Earth's magnetosphere was investigated. Similarly to the XMM-Newton experiment, the data on the X-ray background of the solar corona were carefully selected. Data from large-scale solar events were excluded to obtain pure X-ray background of the solar corona. The coincidence of the energy ranges and the presence of variations in the X-ray background intensity in both experiments cannot be accidental. Most likely, these experiments indicate that, in both cases, there is a common reason for the observed X-ray events.

One can see some contradictions that occur in the course of the long search for one of the candidates for the dark matter particles. On the one hand, the main evidence in favor of the existence of axions is the experimentally proved variations in the X-ray background intensities of the solar corona and the Earth's magnetosphere. On the other hand, axions themselves were not detected in any of the numerous experiments carried out in ground-based laboratories. The situation is strange because there is indirect evidence of the effect, but the axions themselves that are responsible for this effect have not been detected. This work is intended to eliminate this contradiction.

## 2. PHOTON PAIR AS A CANDIDATE FOR THE DARK MATTER PARTICLE. SPLITTING OF PHOTON PAIRS

Photons are the most widespread particles in the Universe. They have a zero rest mass and can propagate at a speed of $c = 300\,000$ km/s. However, such a speed can be achieved only in absolute vacuum, which can be found almost nowhere in nature. In fact, the photon properties commonly postulated in classical physics are a simplified model.

In [11, 12], a generalization of the Gliner–Sakharov hydrodynamic vacuum theory [13–15] was obtained in the framework of the modified general theory of relativity (MGTR), which is in good agreement with the observational data on the accelerated expansion of the Universe. In [11, 12], it was shown that the exact solution to the MGTR equations has a structure corresponding to the quantum theory of gravitation generalized to the case of a finite value of the cosmological constant [16]. From the condition of their exact coincidence, the mass of ultralight scalar bosons continuously produced due to the polarization of vacuum in the course of expansion of the Universe was found to be $m_0 \approx 3 \times 10^{-66}$ g. In this case, the possibility was considered to identify these scalar bosons with photon pairs that are produced from vacuum, have a zero total helicity and (similar to axions) can be dark matter particles.

If we assume that the observed coronal production of photons with energies of $\Delta E \approx 3$ keV is due to the splitting of photon pairs in the magnetic field of the solar corona ($H = 100$ G). Then, from the relation $\Delta E = HM$, we can determine the magnetic moment $M$ of photons in the splitting photon pairs (the volume density of which is determined by the energy density in the corona),

$$M = 4.8 \times 10^{-14} \text{ J/T.} \tag{1}$$

Estimate (1) of the photon magnetic moment can be taken into account when interpreting the observational data from full-scale and laboratory experiments, e.g., experiments on the Faraday effect.

We also note that, in [17], the electric dipole moment associated with each photon pair was also estimated to explain the observed baryon asymmetry (in [17], such photon pairs were called gravitons).

According to the theory proposed in [11], the Universe existed for unlimited time in the past, with no singularities associated with the Big Bang or its inflationary modifications. By this theory, the photon pairs are continuously produced all over the space and the observed microwave background radiation (MBR) should also be a consequence of this process of production of photon pairs from vacuum in the course of expansion of the Universe (which, according to the estimate obtained in [11], should last for another 38 billion years and then change for its compression).

In this case, the MBR can be formed due to the splitting of photon pairs in the interplanetary and intergalactic magnetic field. Indeed, for a typical magnetic field of $H \approx 0.765 \times 10^{-5}$ G and the magnetic moment given by expression (1), the theory yields $T \approx 2.725$ K.

Figure 4 shows a scheme of the photon pair splitting in a magnetic field. The mechanism of this process resembles the Primakov effect, however, without a "mediator" particle (axion). In this context, it is of interest to carry out full-scale and laboratory studies concerning the possibility of photon pair splitting in a magnetic field.

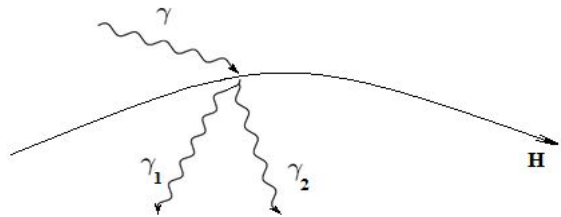

**Fig. 4.** Effect of photon pair splitting in a magnetic field.

In the photon pair under consideration, the two constituent photons (similarly to a single photon) display both the particle and wave properties, depending on their energies and the properties of the ambient medium. The lower the frequency (i.e., the longer the wavelength and the lower the energy), the closer the properties of the photon pair to the wave ones. The higher the frequency (i.e., the shorter the wavelength and the higher the energy), the closer the properties of the photon pair to the particle ones. In this case, it is the photon pair that serves as the quantum of electromagnetic interaction. Figure 5 illustrates a photon pair in terms of the conventional oscillating electric and magnetic field vectors (**E**$_1$ and **E**$_2$, and **H**$_1$ and **H**$_2$, respectively). Each photon in the pair has its own pair of the **E** and **H** vectors, which oscillate at an angle of 90° relative to one another. It is well known that a photon can have either right- or left-handed helicity. Therefore, each photon in this pair can also have right- or left-handed helicity. However, the pair as a whole should have a zero total helicity, i.e., the photon pairs are composed of photons of opposite helicity. It is due to this structure that a photon pair splits in a magnetic field. The properties of a photon pair can be revealed from indirect indications that manifest themselves in the interaction with ambient particles or due to the splitting in a magnetic field.

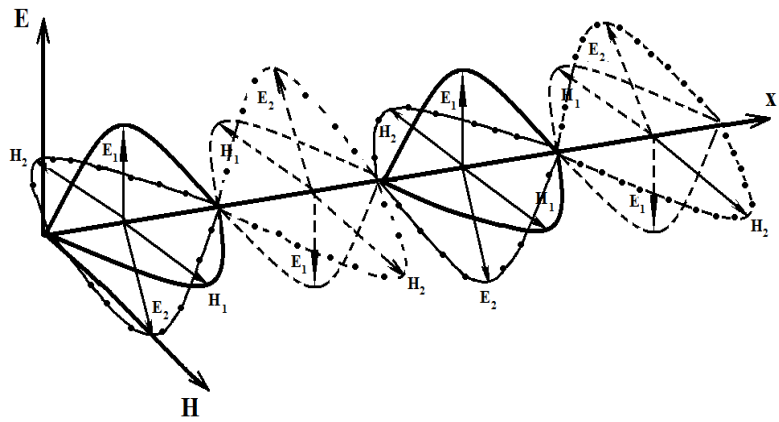

**Fig. 5.** Photon pair.

### 3. MECHANISM OF SOLAR CORONA HEATING

Since photon pairs should split in any magnetic field, such splitting should also occur in the solar corona. It was experimentally ascertained that the spectral intensity of the X-ray background of the solar corona in the photon energy range of 2–6 keV decreases or increases in time. In our opinion, this is a result of the splitting of photon pairs, accompanied by an energy release, which, in turn, results in an increase in the coronal plasma temperature. However, this increase is limited and the temperature of the solar corona plasma is maintained at a constant level of $T = 1.5 \times 10^6$ K. This may be due to the fact that, after splitting, photon pairs can recombine again and also interact with ions and electrons of the hot plasma. A fraction of the energy is radiated from the solar corona into open space, due to which the temperature balance is maintained. Due to the high matter density inside the Sun, the radiation generated in the solar core and propagating through the zone of radiation energy transport toward the solar surface very slowly diffuses through the inner layers of the Sun, because it is permanently absorbed and reradiated by heavy atoms and ions. It is for this reason that the processes of splitting of photon pairs are almost absent inside the Sun and practically do not affect the distribution of temperature, which, as is well known, is maximal in the core and gradually decreases toward the photosphere. When the solar radiation escapes from the photosphere, the matter density drops and the radiation begins to propagate relatively freely. In the chromosphere and, then, in the corona, the flux of photon pairs enters the region with a strong magnetic field, where they split and an energy is released, due to which the coronal plasma is heated to a temperature of $1.5 \times 10^6$ K. Further, different interaction scenarios for the split photon pair can take place: recombination; various types of interactions with ions and electrons of the hot coronal plasma, followed by the emission of secondary photon pairs; etc.

# 4. CONCLUSIONS

Experimental data obtained in the *INTERBALL-Tail* probe, *GOES*, *RHESSI*, and *XMM-Newton* projects made it possible to analyze a number of physical phenomena that occur in the solar corona and Earth's magnetosphere and derive the following estimates and conclusions.

1. The observed variations in the intensity of the X-ray background of the solar corona in the photon energy range of 2–6 keV were confirmed by observations of the X-ray background of the Earth's magnetosphere performed in the XMM-Newton International Space Observatory. A conclusion is made on the common mechanism of these phenomena.

2. To explain the mechanism for variations in the X-ray background intensities in both the solar corona and the Earth's magnetosphere, the quantum theory of photon pairs produced from vacuum of the expanding Universe, as well as in the processes occurring in the solar core, is involved. A hypothesis on the splitting of photon pairs in the interplanetary and intergalactic magnetic fields, as well as in the magnetic field of the solar corona, is proposed.

3. For the energies of photon pairs in the range of 2–6 keV, the photon magnetic moment is estimated for a nonzero mass of ultralight scalar bosons (photon pairs).

4. Assuming that the MBR of the Universe is caused by the splitting of photon pairs and taking into account the magnitudes of the interplanetary and intergalactic magnetic fields and the obtained estimate for the photon magnetic moment, the obtained radiation temperature is found to agree with the well-known MBR temperature $T \approx 2.725$ K.

5. It is hypothesized that the mechanism of coronal plasma heating is due to the slitting of photon pairs in the magnetic field of the solar corona.